\documentclass[12pt]{iopart}

\expandafter\let\csname equation*\endcsname\relax
\expandafter\let\csname endequation*\endcsname\relax

\usepackage{amssymb}
\usepackage{graphicx} 
\usepackage{amsmath}
\usepackage{multirow}
\usepackage{fullpage}
\usepackage{tabularx}
\usepackage{dcolumn}
\usepackage{booktabs}
\usepackage{rotating, makecell}
\usepackage{float}
\usepackage{hyperref}
\usepackage{multicol}
\usepackage{subfig}
\usepackage{derivative}

\begin{document}

\title{Resistivity anomalies and intrinsic spin-orbit coupling in superconducting thin film solid solutions of Nb$_{1-x}$U$_x$ for $0.15 < x < 0.40$}

\author{Syed Akbar Hussain$^1$, Katie Brewer$^1$, Livina Onuegbu$^1$, Lottie M. Harding$^1$, Sean Giblin$^2$, Ross S. Springell$^1$, Christopher Bell$^1$}
\address{$^1$ School of Physics, University of Bristol, Tyndall Avenue, Bristol, BS8 1TL, United Kingdom; $^2$ School of Physics and Astronomy, Queen's Buildings, 5 The Parade, Newport Road, Cardiff, CF24 3AA, United Kingdom}
\ead{christopher.bell@bristol.ac.uk}

\begin{abstract}
Polycrystalline thin films of $\mathrm{Nb}_{1-x}\mathrm{U}_{x}$ solid solutions with  $0.15\leq x \leq 0.40$ were prepared via d.c. magnetron sputtering at ambient conditions. X-ray characterisation of the samples revealed a systematic shift of the (110) Nb Bragg reflection with U concentration, consistent with substitutional replacement of the Nb by U. Superconductivity was observed in all samples below $2$ K. Analysis of the superconducting critical fields revealed a direct scaling of the spin-orbit scattering and transport scattering times, indicating Elliott-Yafet-type spin relaxation in the system. Magnetoresistivity measurements showed a feature in the range $4$ K $\leq T \leq30$ K suggesting a possible a complex interplay between electron-electron interaction and localisation physics.
\end{abstract}


\maketitle

\section{Introduction}
\label{introduction}
The study of electronic transport properties of binary alloys has a long history for both metals and superconductors. The impact of the scattering centres introduced with alloying on the transport properties can be broadly separated into those which induce spin-flips and those which do not. The former includes both magnetic as well as spin-orbit coupling (SOC) effects. SOC has become especially important in contemporary studies of metallic transport due to the interest in phenomena such as the spin Hall effect \cite{sinova_spin_2015, sala_giant_2022}. To enhance the SOC strength, heavy elements such as Pt, Au, W, Ir have been investigated in a variety of alloy systems \cite{kim_enhancement_2020, qu_enhanced_2024, lowitzer_extrinsic_2011}.

In s-wave superconductors, the effect SOC is also  increasingly important area of research, since the simplistic picture of singlet Cooper pairs is no longer valid, and SOC can be leveraged to create novel forms of superconductivity, such as the s-wave odd-frequency triplet state \cite{yang_boosting_2021, bergeret_odd_2005, buzdin_proximity_2005, bergeret_triplet_2023}. So far, most of these studies have focused on layered thin film systems utilising the proximity effect. A key challenge is to maximise the number of odd frequency Cooper pairs, and in this context homogeneous alloys with large SOC are a potential source of increased triplet superfluid density. 

Little is known about the impact of alloying conventional superconductors with heavy actinide elements such as U. Binary alloys of the form  U-T (T = Mo, Pt, Pd, Nb, Zr) have been the subject of a series of studies, most notably using splat cooling \cite{kim-ngan_superconductivity_2018}.U-Nb alloy systems have been studied primarily due to the interest in the stability that body-centred cubic ({\it bcc}) $\gamma-U$ structures provide for nuclear fuels, as well as some studies on radiation damage \cite{ghoshal_creep_2013}. Most of these studies have been carried out in the higher atomic (at$.$) \% U content region ($75-100$ at$.$ \% U). Some studies on the superconductivity have been carried out, but are limited to bulk systems with relatively low Nb content \cite{berlincourt_hall_1959, chandrasekhar_electrical_1958}. 

To the best of our knowledge, no structural or low temperature magnetotransport studies of Nb-U alloys have been performed in the lower at$.$ \%  of U region, where the impact of the SOC from the U on the Nb superconductivity can be studied. Here we focus on this region, tuning the at$.$ \% of U content in Nb-based thin film binary alloys, in order to reveal the impact of the U on superconductivity, and better understand of the interplay between the inherent superconductivity within Nb, the intrinsic SOC coupling from U, and the possible role of the U $5f$-electrons in the system. 

$\mathrm{Nb}_{1-x}\mathrm{U}_{x}$ polycrystalline thin film layers were grown on commercial glass substrates capped with Ge via a d$.$c$.$ magnetron co-sputtering, varying the magnetron power to control the at$.$ \% U content in Nb. Structural characterisation and chemical composition measurements were then made via x-ray diffraction, x-ray reflectivity and spectroscopy techniques. Low temperature  magnetotransport measurements focussed on the anomalous resistivity observed at temperatures in the range $4$ K $\leq T \leq30$ K and the superconductivity below $2$ K.

\section{Experimental Methods}
\label{methods}
All samples in this study were grown using the dedicated actinide d$.$c$.$ magnetron sputtering facility at the University of Bristol, UK \cite{farms, springell2022review}. This ultra-high vacuum system has a base pressure $\sim 10^{-10}$ mbar and contains four sputtering guns inside a load-locked chamber. Samples were fabricated by co-sputtering from separate (depleted) U and Nb targets and capped with Ge. Sputtering took place in an Ar atmosphere of $7.3\times10^{-3}$ mbar. Deposition rates for U, Nb and Ge were individually calibrated by ex-situ x-ray reflectivity measurements. Typical deposition rates were $0.1 - 0.6$ \AA/s, and were tuned by varying the sputtering power in the range of $5 - 10$ W for U and $50 - 70$ W for Nb. The substrates were kept at ambient temperature during growth. Film thicknesses were nominally $\sim$ between $40-50$ nm for all four samples. 

 The substrates were $10$ mm $\times$ $10$ mm $\times$ $0.7$ mm Corning EAGLE XG glass substrates, polished to optical grade.  {\it Ex-situ} x-ray diffraction (XRD) and x-ray reflectivity (XRR) experiments were carried out using an in-house four-axis Panalytical X'pert diffractometer in the Bragg-Brentano geometry. The measurements were carried out at room temperature with a Cu-K$_{\alpha}$ source. The XRR data were analysed using GenX \cite{bjorck_genx_2007, glavic_genx_2022}, to obtain thickness, atomic density, surface and interface roughness. XRD measurements were performed with an angular step size of $\Delta2\theta = 0.02^{\circ}$. 

Sample composition was estimated using energy dispersive x-ray spectroscopy (EDXS) in a scanning electron microscope (Oxford Instruments UltimMax EDXS and analysed with the AZtech software) using beam voltage of $20$ kV. Six spectra we averaged, with three taken at two separate sites. Table \ref{tab:Sample growth summary} summarises the samples fabricated in this study, together with the at$.$ \% of U estimated from the EDXS data. These estimates will be used in the rest of the paper when referring to the sample composition. Two separate resistance versus temperature (RT) measurements were carried out. From room temperature to 4.2 K a dip probe was utilised with the samples electrically bonded in a Van der Pauw configuation (the samples were cut into $5$ mm $\times$ $5$ mm squares using a diamond wire saw for this purpose). The samples were then further cut into $1$ mm $\times$ $5$ mm bars for magnetotransport studies down to $0.4-0.5$ K at applied fields between $0-2250$ mT in a Quantum design PPMS system with He-3 option. The applied magnetic field in the PPMS was applied parallel to the surface of the samples, perpendicular to the bias current.

 \begin{table}[!h]
     \centering
     \begin{tabular}{|c|c|c|c|c|c|}
     \hline
     &  \multicolumn{3}{c|} {Target power (W)} & Atomic \% U & Film thickness, $d_{\mathrm{Nb:U}}$ \\ 
     \cline{2-4} 
     Sample number &  Nb  & U  & Ge  & (at. \% U) & (nm) \\
     \hline
         SN2488 & 70 & 5 & 20 & $15.8 \pm 0.3$ & $47.1$ \\
    \hline
         SN2487 & 60 & 5 & 20  & $19.3 \pm 0.1$ & $39.8$ \\
    \hline
         SN2486 & 60 & 10 & 20 & $31.5 \pm 0.2$ & $52.4$\\
    \hline
         SN2485 & 50 & 10 & 20  & $39.0 \pm 0.6$ & $46.4$ \\
    \hline
     \end{tabular}
     \caption{Summary of the sputtering growth parameters, the at$.$ \% of U in the samples estimated by EDXS and the estimated thickness of the alloy layer, $d_{\mathrm{Nb:U}}$ via XRR fit from GenX.}
     \label{tab:Sample growth summary}
\end{table}

\section{Results}

\subsection{X-ray diffraction and characterisation}
 
\begin{figure}[h]
    \centering
    \includegraphics[width=1.0\linewidth]{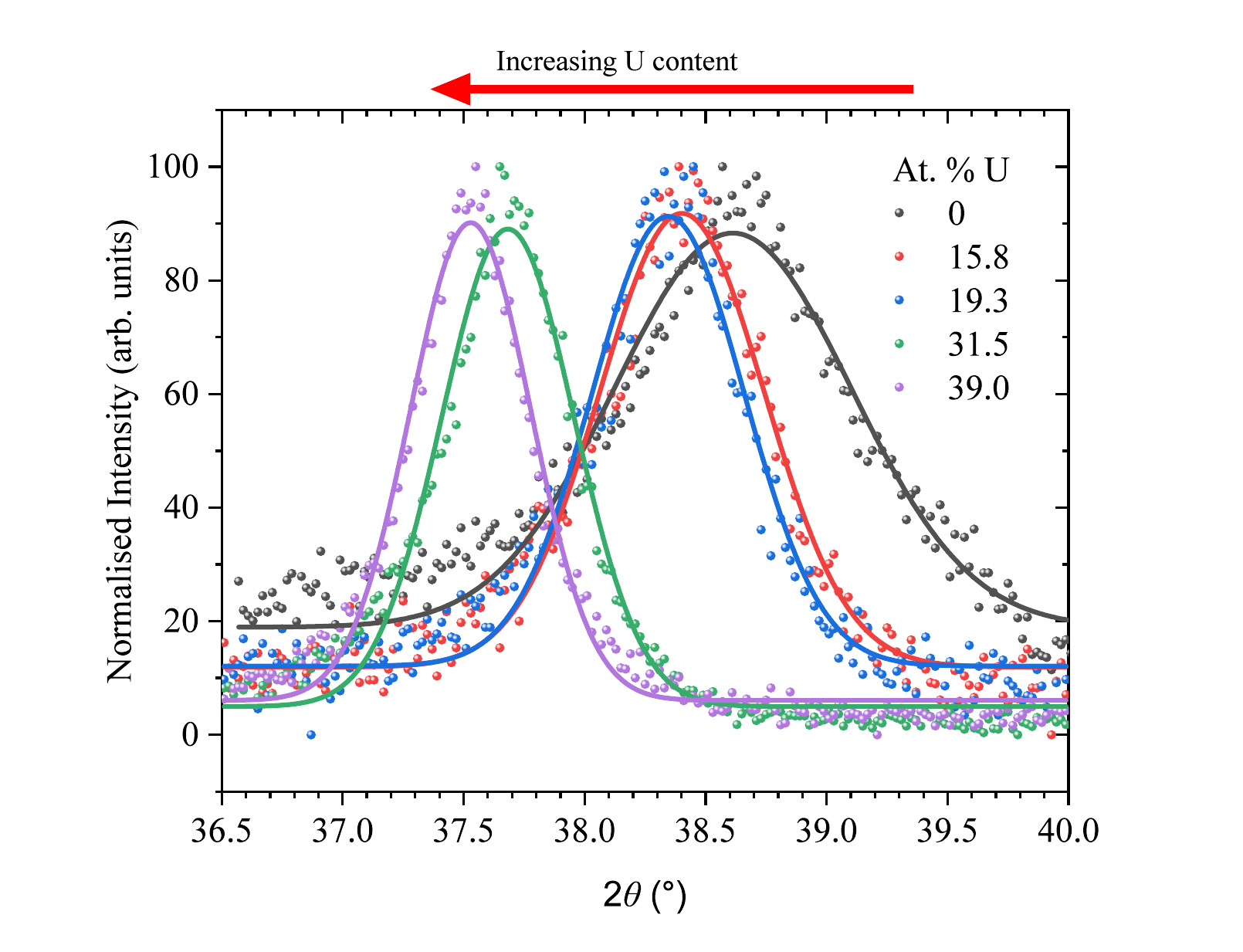}
    \caption{Intensity versus $2\theta$ around the (110) peak of Nb for the four samples and a Nb reference. The solid lines shown are Gaussian fits made to the respective data sets. The peak positions clearly shift to lower angles with increasing U concentration.  A reduction of the FWHM of the peak with increasing U concentration is also observed.}
    \label{fig:HA}
\end{figure}

High angle XRD data from $2\theta-\omega$ scans of each sample are shown in Figure \ref{fig:HA}, focussing on the angles around the $(110)$ Bragg reflection of {\it bcc} Nb.  A measurement of the (110) position of a pure Nb sample grown within the same chamber and conditions has also been included as a reference. The (110) Bragg peak of Nb is seen to shift to lower angles with increasing U content, indicating an increasing lattice constant, $a_{\mathrm{xrd}}$. A Gaussian fitting was performed to estimate both the peak positions, as well as the full width at half-maximum (FWHM).  The FWHM decreases as the U content in the system increases, suggesting a possible increase in crystallite size. 

\begin{figure}[h]
    \centering
    \includegraphics[width=1.0\linewidth]{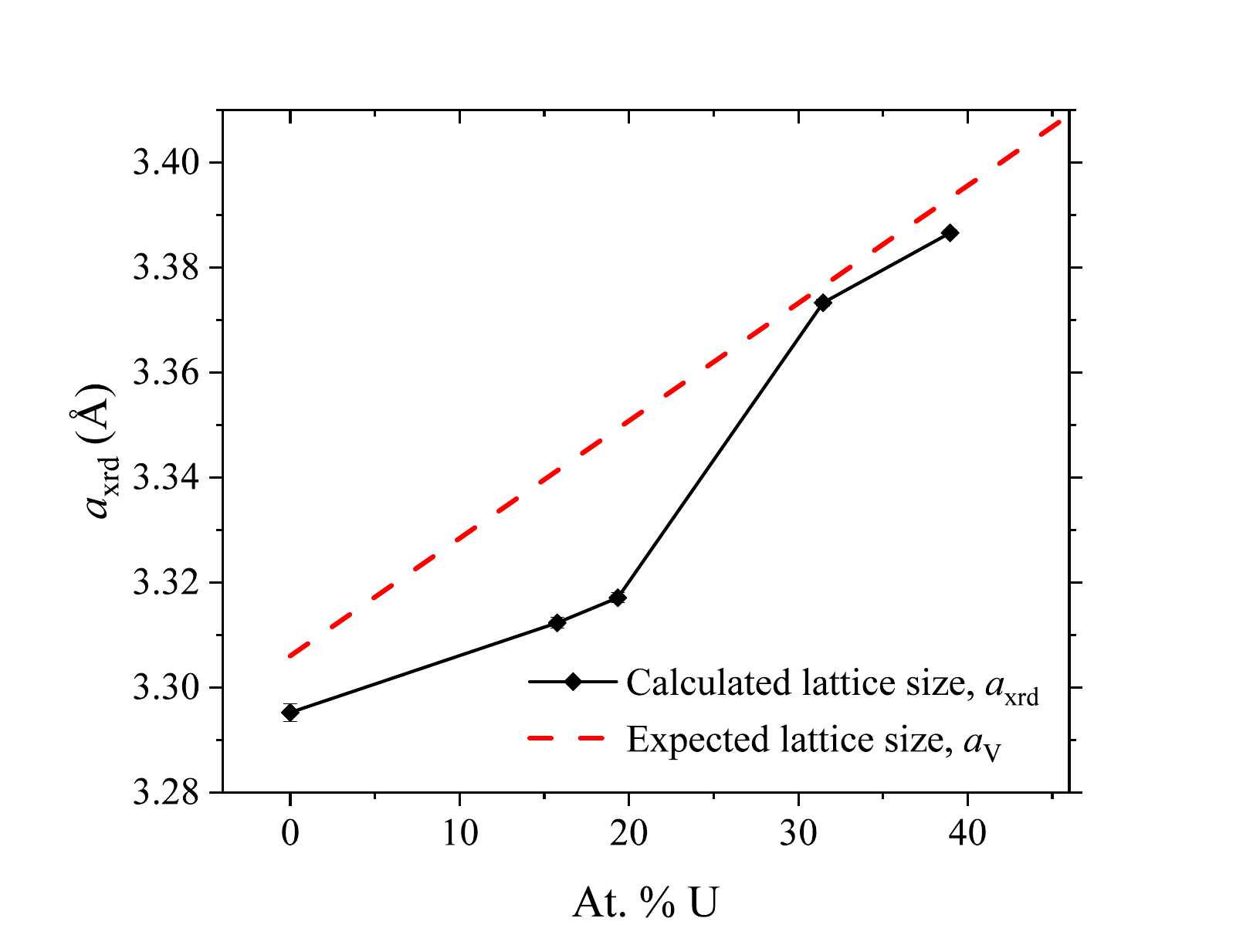}
    \caption{Lattice constant, $a_{\mathrm{xrd}}$ versus at$.$ \% U. $a_{\mathrm{xrd}}$ was calculated from the peak of the Gaussian fit made from the $2\theta - \omega$ scan of the (110) Nb Bragg peak. The dashed line represents the expected variation in the lattice constant, $a_{\mathrm{V}}$, calculated from Vegard's law using values of $a_{\mathrm{Nb}} = 3.306$ \AA \cite{roberge_lattice_1975} and $a_{\mathrm{U}}=3.53$ \AA \cite{chakraborty_micro-structural_2015}. Other lines are guides to the eye. Error bars on the $a_{\mathrm{xrd}}$ data are small compared to the point size. }
    \label{fig:VL}
\end{figure}

The lattice constant calculated from the XRD measurement shown in Figure \ref{fig:VL} is seen to increase with at$.$ \% of U. The dashed line shown serves as a reference of the expected lattice constant, $a_{\mathrm{V}}$ that was calculated using Vegard's Law, $a_{\mathrm{V}} = a_{\mathrm{Nb}}(1-x) + a_{\mathrm{U}}x $, where $x$ is the atomic fraction of U, and $a_{\mathrm{Nb}}$ and $a_{\mathrm{U}}$ are the literature lattice constants for Nb and U, respectively. Although the lattice parameter evaluated from XRD is systematically slightly lower than the literature values, the trend of increasing lattice parameter agrees well with the expected trend from Vegard's law, suggesting that the U atoms are entering the Nb {\it bcc} structure substitutionally. 

\begin{figure}[h]
    \centering
    \includegraphics[width=1.0\linewidth]{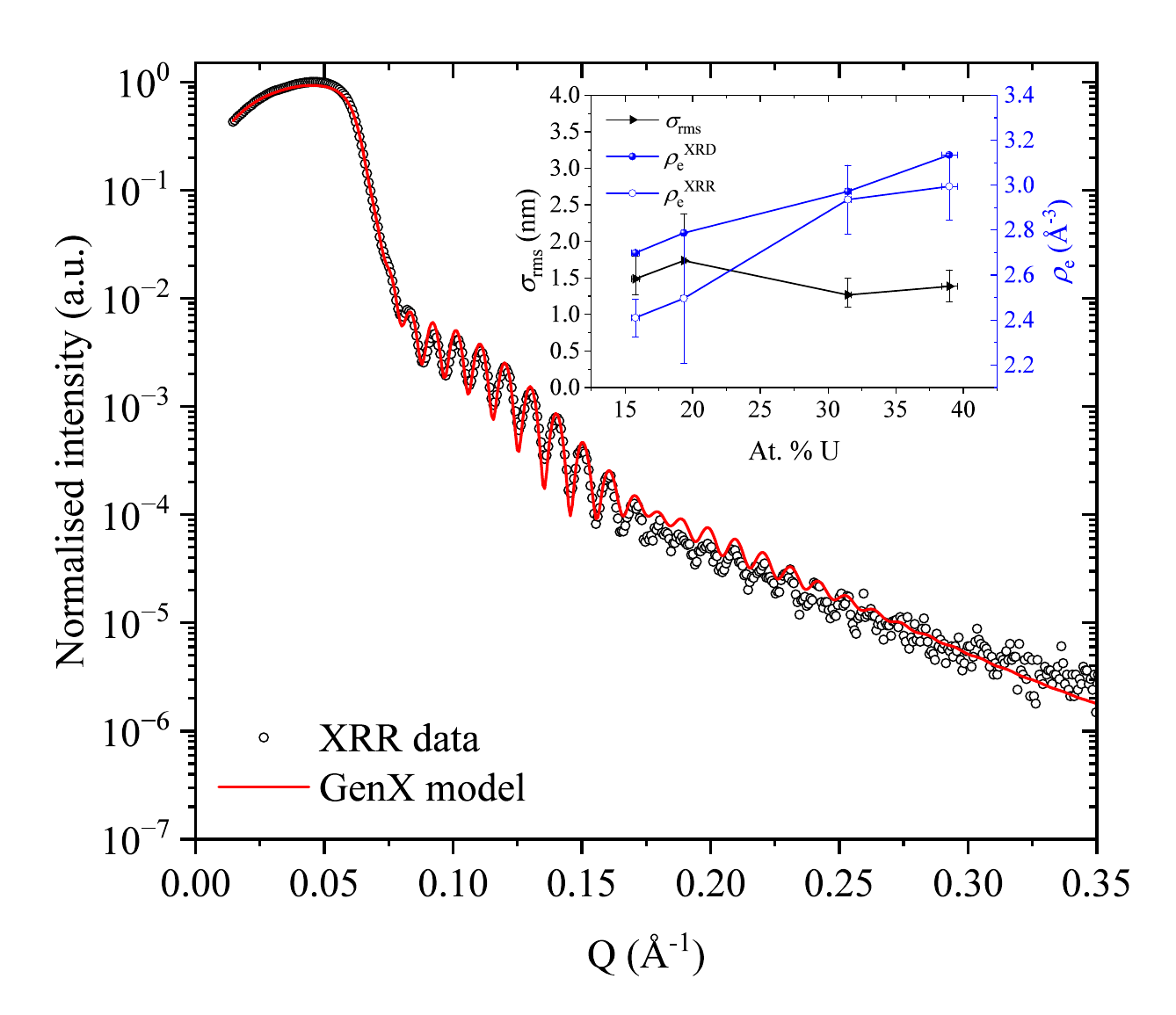}
    \caption{Fit to the reflectivity data for the sample with U composition = 19.3 at$.$ \% U. Data are open symbols and line is the best fit. Inset: Root mean square roughness, $\sigma_{\mathrm{rms}}$, (triangle symbols) versus composition of the Nb:U alloy layer extracted from the reflectivity fit. Variation with composition  of the electron density, extracted from XRD data, $\rho_{\mathrm{e}}^{\mathrm{XRD}}$,  and from the XRR data, $\rho_{\mathrm{e}}^{\mathrm{XRR}}$ are also shown as closed and open circles, respectively.}
    \label{fig:XRR-FIT}
\end{figure}
An example of the XRR data obtained on one of the samples is shown in Figure \ref{fig:XRR-FIT}, together with the fit made using the GenX software. Excellent agreement between the model and the data allows us to extract various parameters with high confidence. Firstly, the inset of Figure \ref{fig:XRR-FIT} shows the root mean square (rms) roughness of the Nb:U layer, $\sigma_{\mathrm{rms}}$,  for all four samples to be around $\sim 1.5$ nm independent of composition, indicating good quality smooth layers. The inset of the plot also shows the variation with composition of the electron density, $\rho_{\mathrm{e}}^{\mathrm{XRR}}$, determined from the XRR fit. As expected this increases systematically with $x$. Here we used the EDXS compositions to define the weightings of the scattering length density for the Nb:U layer. The $\sigma_{\mathrm{rms}}$ of the Ge cap was determined to be nominally $\sim 1$ nm.

Another measure of the electron density can be determined from the XRD measurement $\rho_{\mathrm{e}}^{\mathrm{XRD}}$. $\rho_{\mathrm{e}}^{\mathrm{XRD}}$ was determined by assuming that a {\it bcc} unit cell always contains 2 atoms per unit cell, with an effective electron number intermediate between 41 (Nb) and 92 (U) determined by the composition fraction $x$. Hence, by using the $x$ value from the EDXS data, and the $a_{\mathrm{xrd}}$ values from the XRD data (Figure \ref{fig:VL}), $\rho_{\mathrm{e}}^{\mathrm{XRD}}$ can be calculated. The increasing trend of $\rho_{\mathrm{e}}^{\mathrm{XRD}}$ tracks closely the $\rho_{\mathrm{e}}^{\mathrm{XRR}}$ variation, and can be understood by noting that for a given increase in effective number of U atoms in a unit cell, the relative increase in the unit cell volume is small compared to the increase in electron density. 

\subsection{Electronic and magnetotransport measurments}
\begin{figure}[h]
    \centering
    \includegraphics[width=1.0\linewidth]{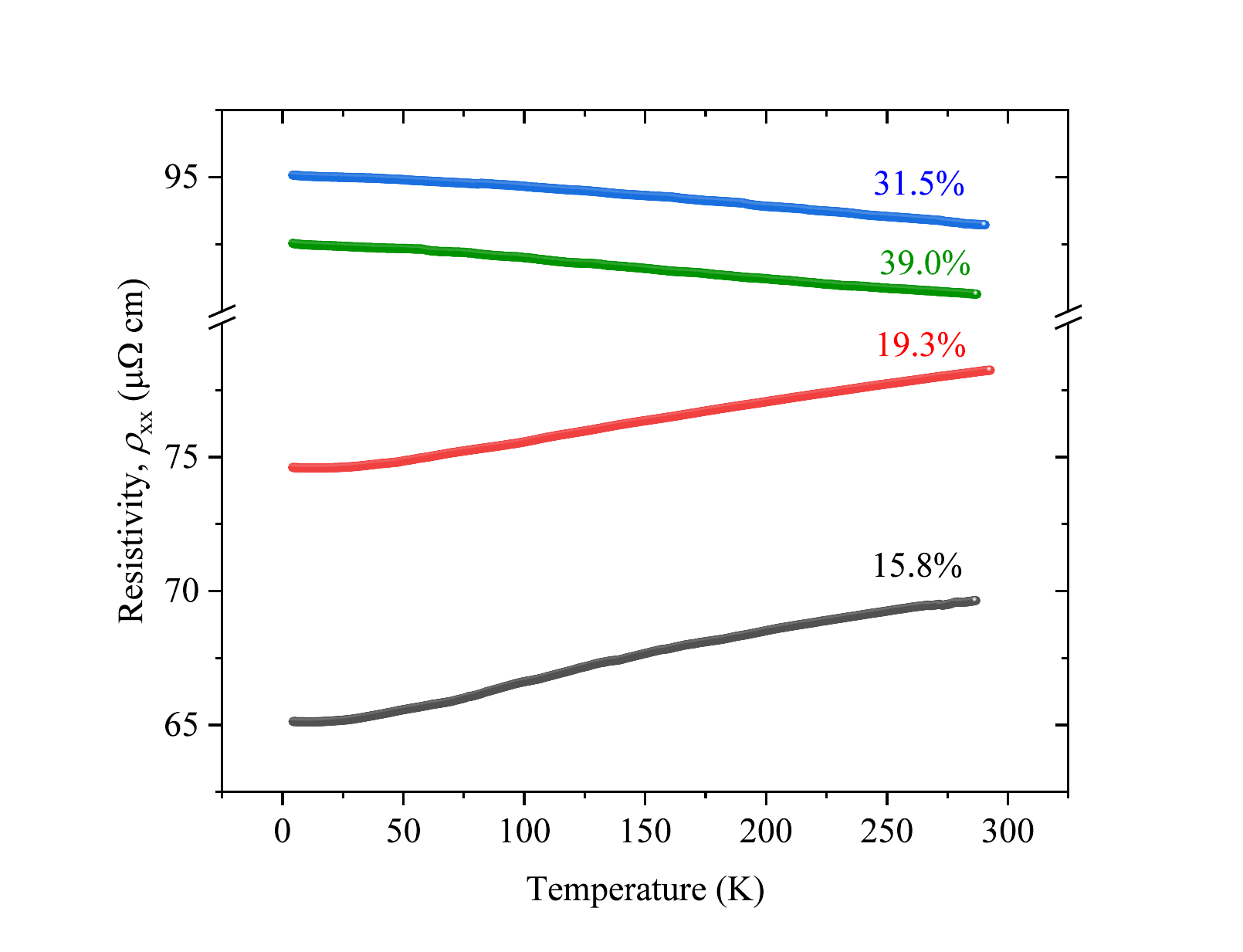}
    \caption{$\rho_{\mathrm{xx}}(T)$ for $4.2$ K $< T < 300$ K  measured in a Van der Pauw configuration. Composition at$.$ \% U content determined from the EDXS data.}
    \label{fig:RTPlotDipper}
\end{figure}

\begin{figure}[h]
    \centering
    \includegraphics[width=1.0\linewidth]{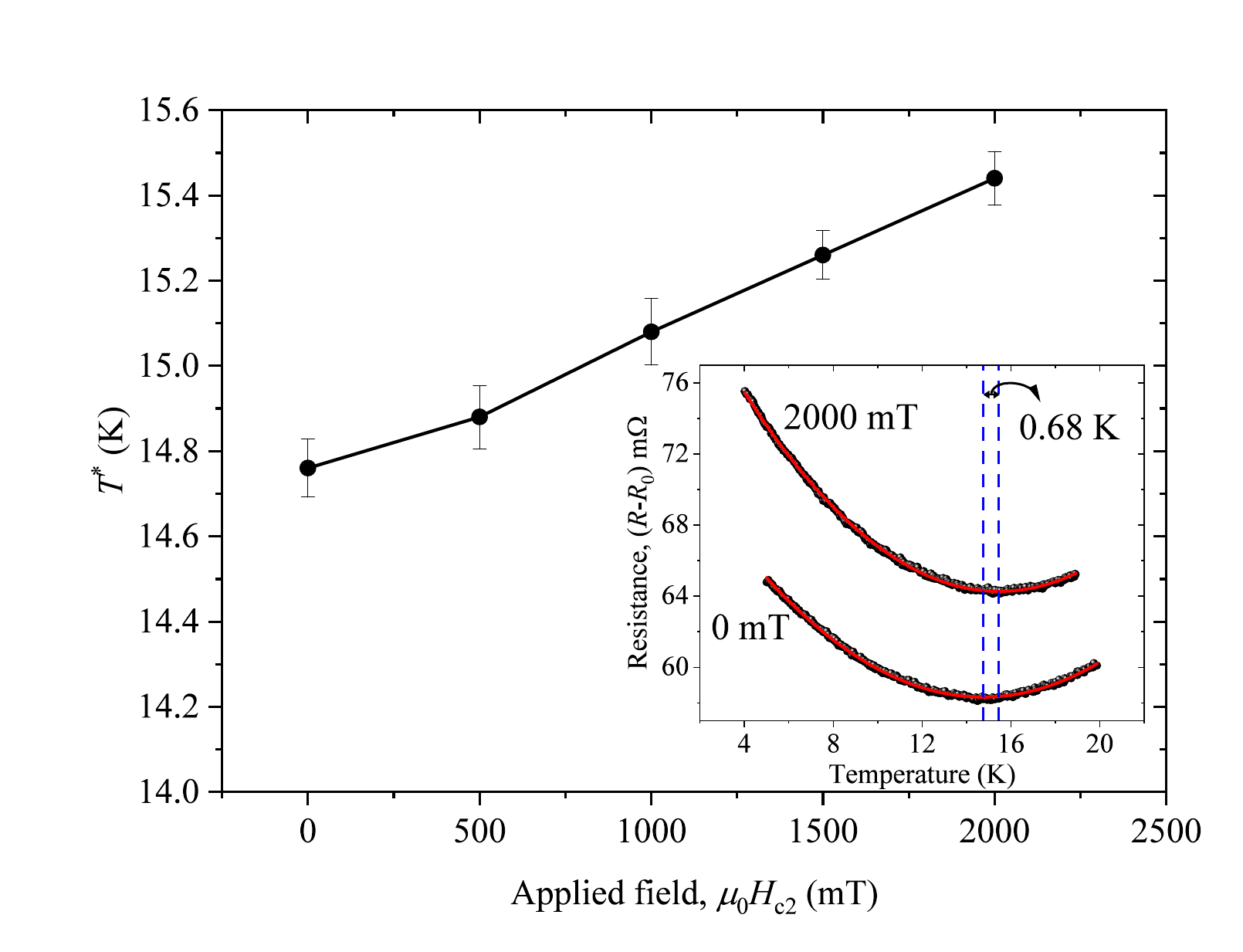}
    \caption{ Magnetic field dependence of $T^{\star}$. Line is a guide to the eye. Inset: Points are the measurements of resistance, $R$, for the sample with at$.$ \% U = 15.8 with the lowest and highest applied in-plane field, as shown. Here an offset of $R_{\mathrm{0}} = 14.6$ $\Omega$  has been applied. Lines are the parabolic fits to the determine the position of $T^{\star}$, the minimum of the parabolas, indicated by the vertical dashed lines.}
    \label{fig:MinimaShift}
\end{figure}

Having discussed the structural aspects of the samples, we next focus on the transport properties of the films. The temperature dependence of the longitudinal of resistivity, $\rho_{\mathrm{xx}}(T)$ for each sample down to $T = 4.2$ K is shown in Figure \ref{fig:RTPlotDipper}. The functional forms of the $\rho_{\mathrm{xx}}(T)$ transitioned from typical metallic character ($\mathrm{d} \rho_{\mathrm{xx}}(T) / \mathrm{d}T > 0)$ to a more semiconducting behaviour for the two higher at$.$ \% samples, although even for these samples the resistivity is below the Mooij limit at low temperatures.
Additionally, in detail, a small feature in the $\rho_{\mathrm{xx}}(T)$ data was clear for the two metallic samples in the range $4$ K $\leq T \leq 20$ K. This temperature at which this feature in the $\rho_{\mathrm{xx}}(T)$ occurs, $T^{\star}$,  was determined via a local parabolic fit.  Magnetotransport measurement within this temperature region revealed a small, but systematic shift of $T^{\star}$ with applied field as shown in Figure \ref{fig:MinimaShift} on the sample with $15.8$ at$.$ \% U. The possible origins of these phenomena are further discussed in the next section.

\begin{figure}[h]
    \centering
    \includegraphics[width=1.0\linewidth]{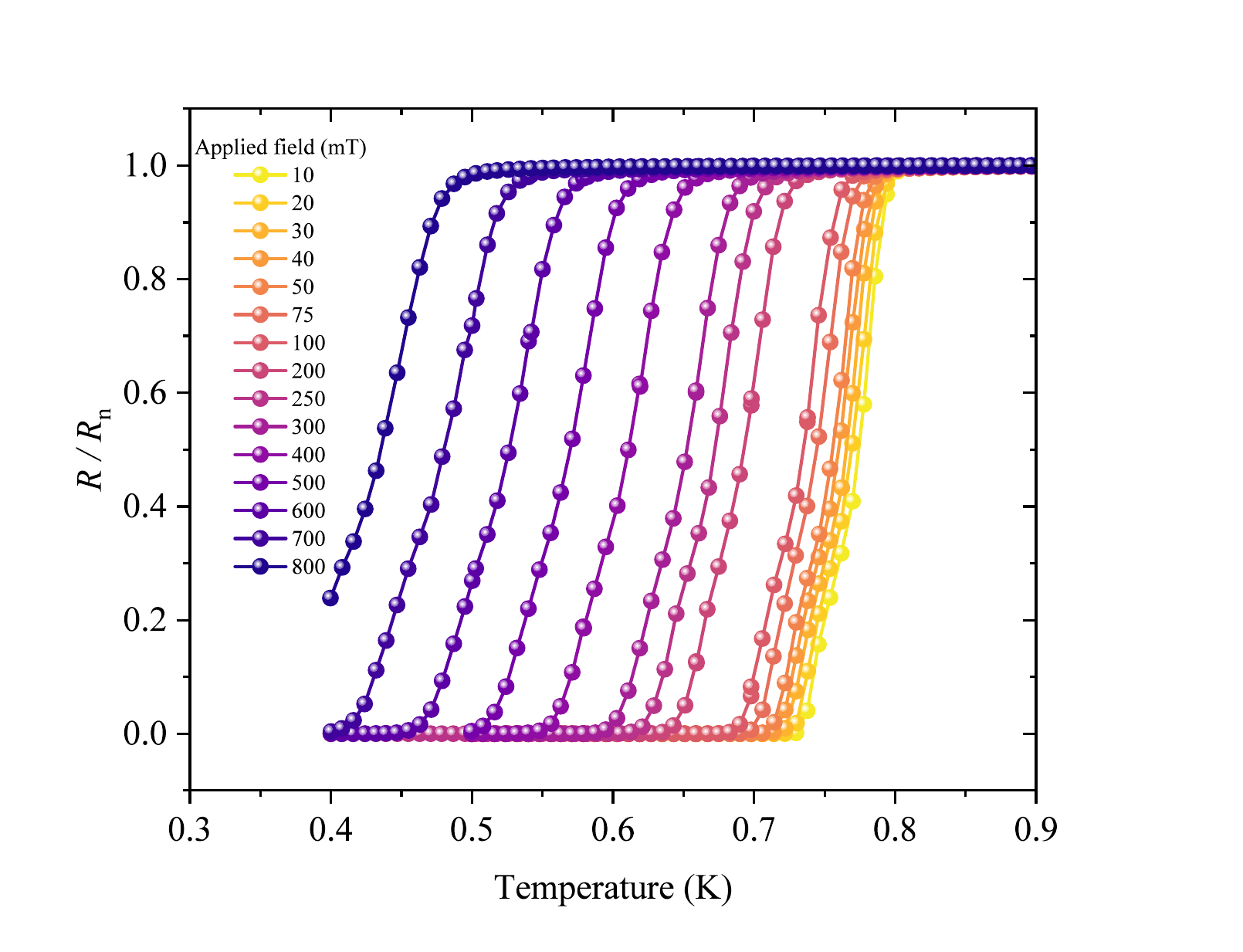}
    \caption{Normalised resistance, $R$ versus $T$ at various applied magnetic fields where $R_{\mathrm{n}}$ is the residual resistance at T = 0.9 K. Sample has at$.$ \% U$ = 31.5$. A full  superconducting transition can be seen for fields up to $\mu_{0}H = 800$mT down to a base $T = 0.4$ K. }
    \label{fig:Field-Measurement}
\end{figure}

Magnetotransport measurements down to $\sim 0.4 $ K were carried out focussing on the superconducting region. Typical $R(T)$ data for one sample (with $19.3$ at$.$ \% U) are shown in Figure \ref{fig:Field-Measurement}. As expected we see a systematic shift to lower temperatures of the superconducting transition temperature, $T_{\mathrm{c}}$, at higher applied magnetic fields, together with a general broadening of the superconducting transition. An estimate of $T_{\mathrm{c}}$ of these samples was extracted from the $R(T)$ data by fitting a sigmoid function $S(x) = L\left[ 1+\e^{-\frac{(x-x_0)}{\sigma}}\right]^{-1}$ close to the transition. Here $L$ is the amplitude of the sigmoid function taken as the values of the residual resistivity, $R_\mathrm{n}$, $x_0$ is the midpoint of the curved, $L/2$ and $\sigma$ controls the steepness of the curve. The value of $T_{\mathrm{c}}$ was then defined the temperature at which the resistance, $R$ of the sample was $50\%$ of the residual resistance value, $R_{\mathrm{n}}$. To provide a visual estimate of the width of the transition the error bars on $T_{\mathrm{c}}$ were determined from the $90\%$ and $10\%$ values of $R_{\mathrm{n}}$. 

\begin{figure}[h]
    \centering
    \includegraphics[width=1.0\linewidth]{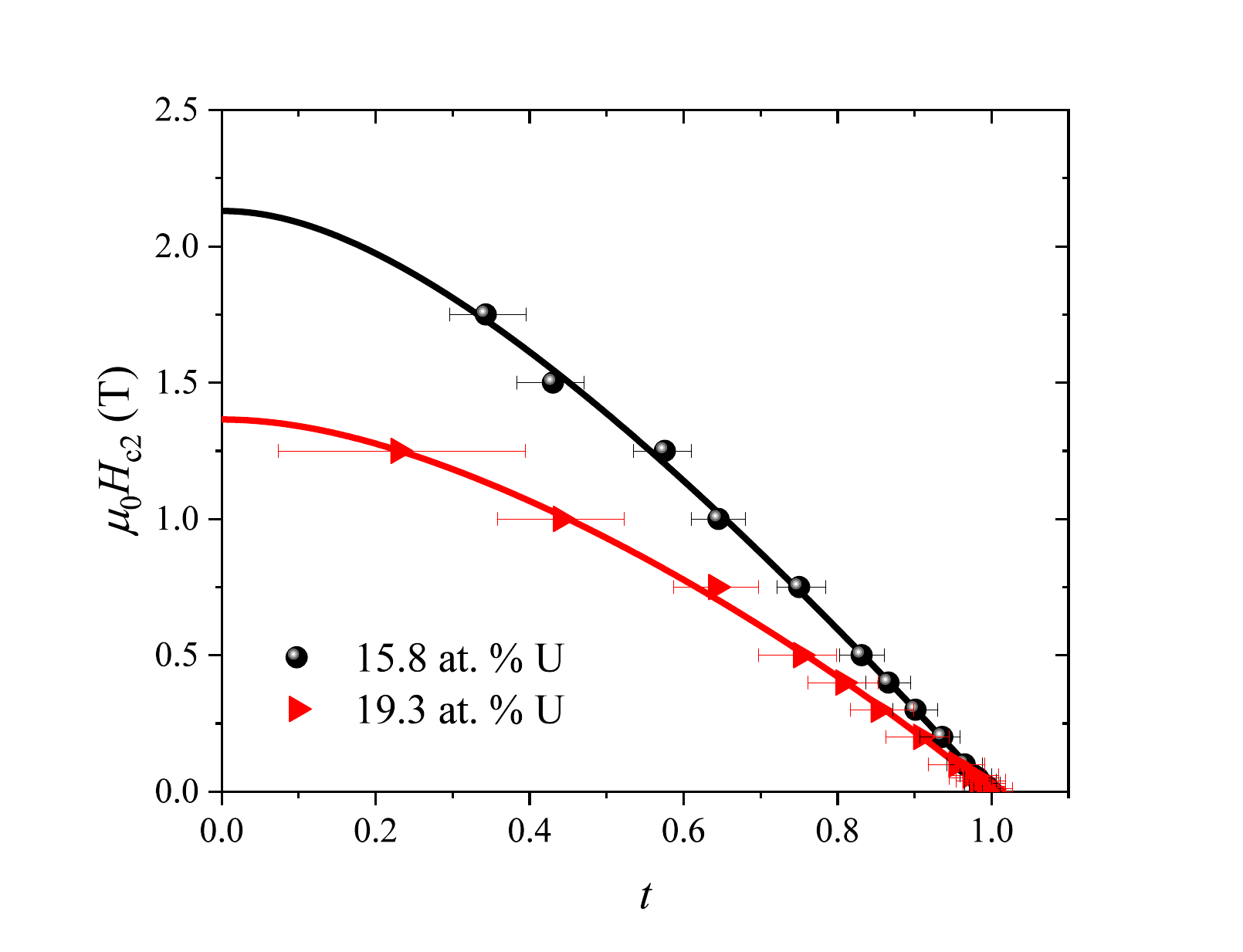}
    \caption{WHH fits (solid lines) to the $H_{\mathrm{c2}}(t)$ data for the samples with $15.8$ (circles) and $19.3$ (triangles) at$.$ \% U., respectively. }
    \label{fig:WHH88}
\end{figure}

\begin{figure}[h]
    \centering
    \includegraphics[width=1.0\linewidth]{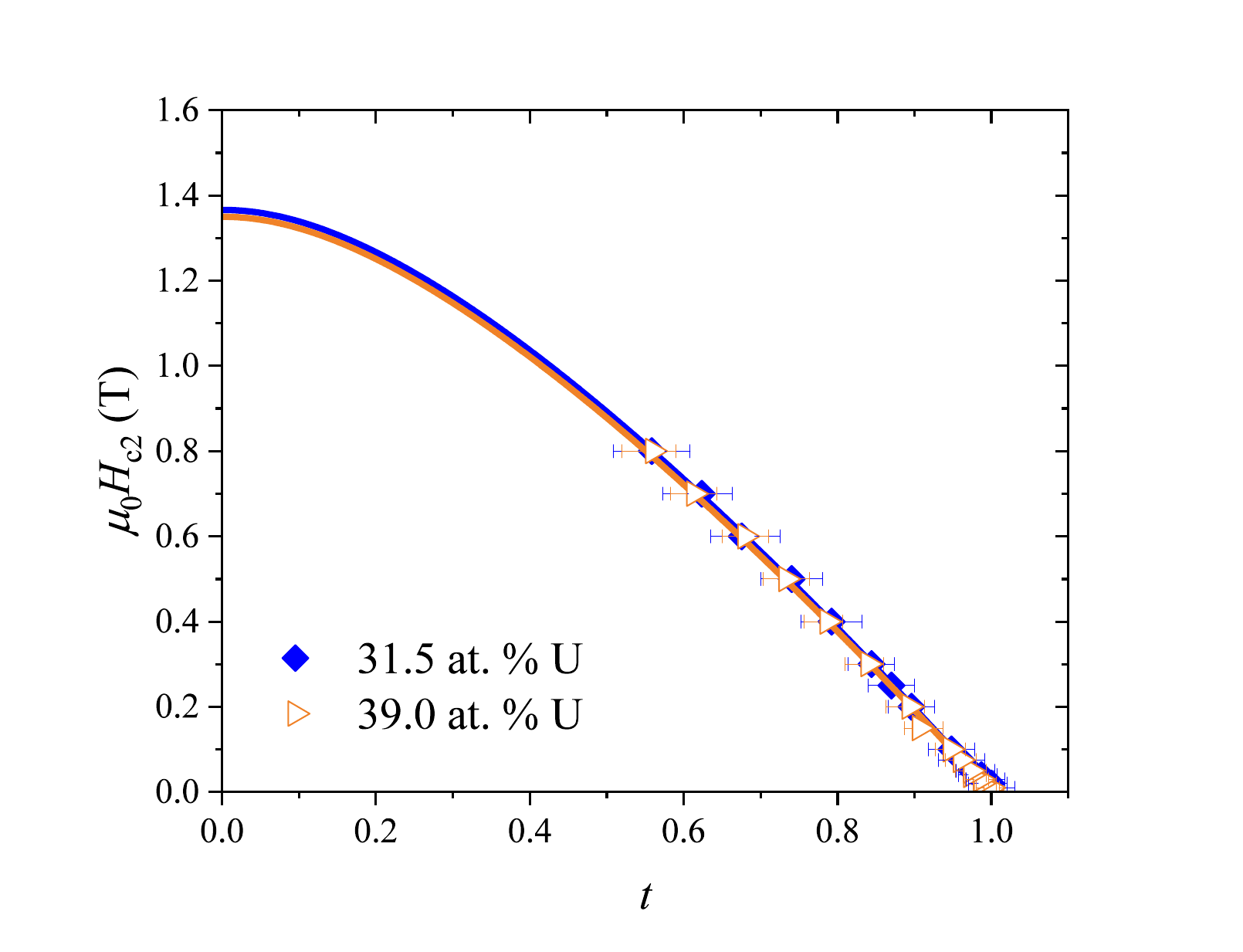}
    \caption{WHH fits (solid lines) to the $H_{\mathrm{c2}}(t)$ data for the samples with $31.5$ (diamonds) and $39.0$ (open triangles) at$.$ \% U. }
        \label{fig:WHH86}
\end{figure}

To further analyse the critical field behaviour, the Werthamer, Helfand and Hohenberg (WHH) theory \cite{werthamer_temperature_1966, mayoh_superconductivity_2019} was used to fit the $H_{\mathrm{c2}}(T)$ data. We use the full form given by
\begin{equation}
    \ln\Bigg(\frac{1}{t}\Bigg) = \Bigg(\frac{1}{2} + \frac{i\lambda_{\mathrm{so}}}{4\gamma}\Bigg)\psi\Bigg(\frac{1}{2}+\frac{h+\frac{1}{2}\lambda_{\mathrm{so}}+i\gamma}{2t}\Bigg)+\\
      \Bigg(\frac{1}{2} - \frac{i\lambda_{so}}{4\gamma}\Bigg)\psi\Bigg(\frac{1}{2}+\frac{h+\frac{1}{2}\lambda_{\mathrm{so}}+i\gamma}{2t}\Bigg) - \psi\Bigg(\frac{1}{2}\Bigg) \; , \label{WHHEqn}
\end{equation}
where $t = T/T_{\mathrm{c}}$, $ h = \frac{4H_{\mathrm{c2}}}{\pi^2}\Big(\frac{dH_{c2}}{dt}\Big)^{-1}\Bigr\vert_{t=1}$, $\gamma = \sqrt{(\alpha_{\mathrm{M}}h)^2 - (\frac{\lambda_{\mathrm{so}}}{2})^2}$ and $\psi$ is the digamma function. Additionally $\lambda_{\mathrm{so}}$ is the spin-orbit scattering (SOS) parameter, and $\alpha_{\mathrm{M}}$ is the Maki parameter \cite{maki_magnetic_1964}. The fit implemented a genetic algorithm that uses a number of possible solutions within the sample space to test the fitness of the solutions, followed by the creation of a next set of possible solutions via selection, recombination and mutation to add randomness \cite{sivanandam_genetic_2008, mccall_genetic_2005, godfrey_polygamma_2024}. The resultant fits to Eq$.$ \ref{WHHEqn} for the four samples are shown in Figures \ref{fig:WHH88} and \ref{fig:WHH86}. We find good agreement with the model for all four samples, with the best fit parameters listed in Table \ref{tab:Hc-table-summary}. 

\begin{table}[!h]
    \centering
    \begin{tabular}{|c|c|c|c|c|c|c|c|}
    \hline
     at$.$ \% U    & $T_c$ $(H_{\mathrm{c2}} = 0)$& $\mu_{0}H_{c2}^{\mathrm{WHH}}$ &  $\mu_{0}H_{\mathrm{P}}^{\mathrm{BCS}}$ & $H_{\mathrm{P}}^{\mathrm{BCS}}/H_{c2}^{\mathrm{WHH}}$  & $\lambda_{\mathrm{so}}$ &    \multirow{2}{2em}{$\alpha_{\mathrm{M}}$} & $l_{\mathrm{tr}}$\\
     & (K) & (T) &  (T) &    &  &   & (\AA) \\
     \hline
     15.8 & $1.72$ &  $2.13$ & $3.20$ &  $1.50$  & $19.43$   & $0.94$ & $7.41$\\
     \hline
     
     19.3 & $1.09$  &  $1.36$ &  $2.04$ & $1.50$ &  $46.31$  & $1.09$ & $5.51$\\
     \hline 
     31.5 & $0.775$ &  $1.37$ & $1.44$  & $1.05$ & $83.31$  & $1.35$ & $4.80$\\
     \hline
     
     39.0 & $0.749$ & $1.35$  &  $1.40$ & $1.03$ &  $99.86$  & $1.37$ & $4.69$\\
     \hline
     
    \end{tabular}
    \caption{Extracted parameters from the WHH fits to the $H_{\mathrm{c2}}(t)$ data. Critical temperature, $T_{\mathrm{c}}$ at $H = 0$ and the upper critical field $\mu_{0}H_{\mathrm{c2}}^{\mathrm{WHH}}$ were extracted from the fits, the latter extrapolated to $T=0$ K. The BCS single band, singlet, Pauli paramagnetic limit, was calculated via $\mu_0 H_{\mathrm{P}}^{\mathrm{BCS}} = 1.84 T_{\mathrm{c}}$. The values of $\lambda_{\mathrm{so}}$ and $\alpha_{\mathrm{M}}$ are from the WHH fit, and the mean free path $l_{\mathrm{tr}} = v_{\mathrm{F}}\tau_{\mathrm{tr}}^{\mathrm{WHH}}$ and $v_{\mathrm{F}} = 1.37\times10^{6}$ m/s which is the bulk Fermi velocity of Nb \cite{ashcroft_solid_2011}.   }
    \label{tab:Hc-table-summary}
\end{table}

While there is a clear general trend of decreasing $H_{\mathrm{c2}}(0)$ and $T_{\mathrm{c}}$ with increasing at$.$ \% U, and all four samples have a $H_{\mathrm{c2}}(0)$ value below the Pauli limiting field $H_{\mathrm{P}}^{\mathrm{BCS}}$, the ratio $H_{\mathrm{P}}^{\mathrm{BCS}}/H_{\mathrm{c2}}(0)$ systematically decreases from $\sim 1.5$ to $\sim 1.05$ from the lowest to highest at$.$ \% U. Concomitantly $\alpha_{\mathrm{M}} = \sqrt{2} H_{\mathrm{c2}}^{\mathrm{orb}}/H_{\mathrm{P}}^{\mathrm{BCS}}$ increased with increasing of at$.$  \% U. The BCS Pauli limit, given in Tesla as $\mu_{0}H_{\mathrm{P}}^{\mathrm{BCS}} = 1.84T_{\mathrm{c}}$  and the dirty limit orbital upper critical field, $\mu_{0}H_{\mathrm{c2}}^{\mathrm{orb}}$ were calculated using the relation $H_{\mathrm{c2}}^{\mathrm{orb}} = -0.69 T_{\mathrm{c}}\frac{dH_{\mathrm{c2}}}{dT}\Big|_{T_{c}}$.

In order to better understand scattering in the system, two key parameters can be extracted from the WHH theory, namely the transport scattering (TRS) time, $\tau_{\mathrm{tr}}^{\mathrm{WHH}}$, given by $\tau_{\mathrm{tr}}^{\mathrm{WHH}}= -12 (v_{\mathrm{F}}^2 e\pi)^{-1}k_{\mathrm{B}}T_{\mathrm{c}} \frac{dH_{\mathrm{c2}}}{dT}\Big|_{T_\mathrm{c}}$, with $v_{\mathrm{F}}$ the Fermi velocity, and the SOS time, $\tau_{\mathrm{so}}^{\mathrm{WHH}}$ \cite{kim_intrinsic_2012} calculated using the relationship $\tau_{\mathrm{so}}^{\mathrm{WHH}} = 2\hbar(3\pi k_{\mathrm{B}}T_{\mathrm{c}}\lambda_{\mathrm{so}})^{-1}$. 

\begin{figure}[h]
    \centering
    \includegraphics[width=1.0\linewidth]{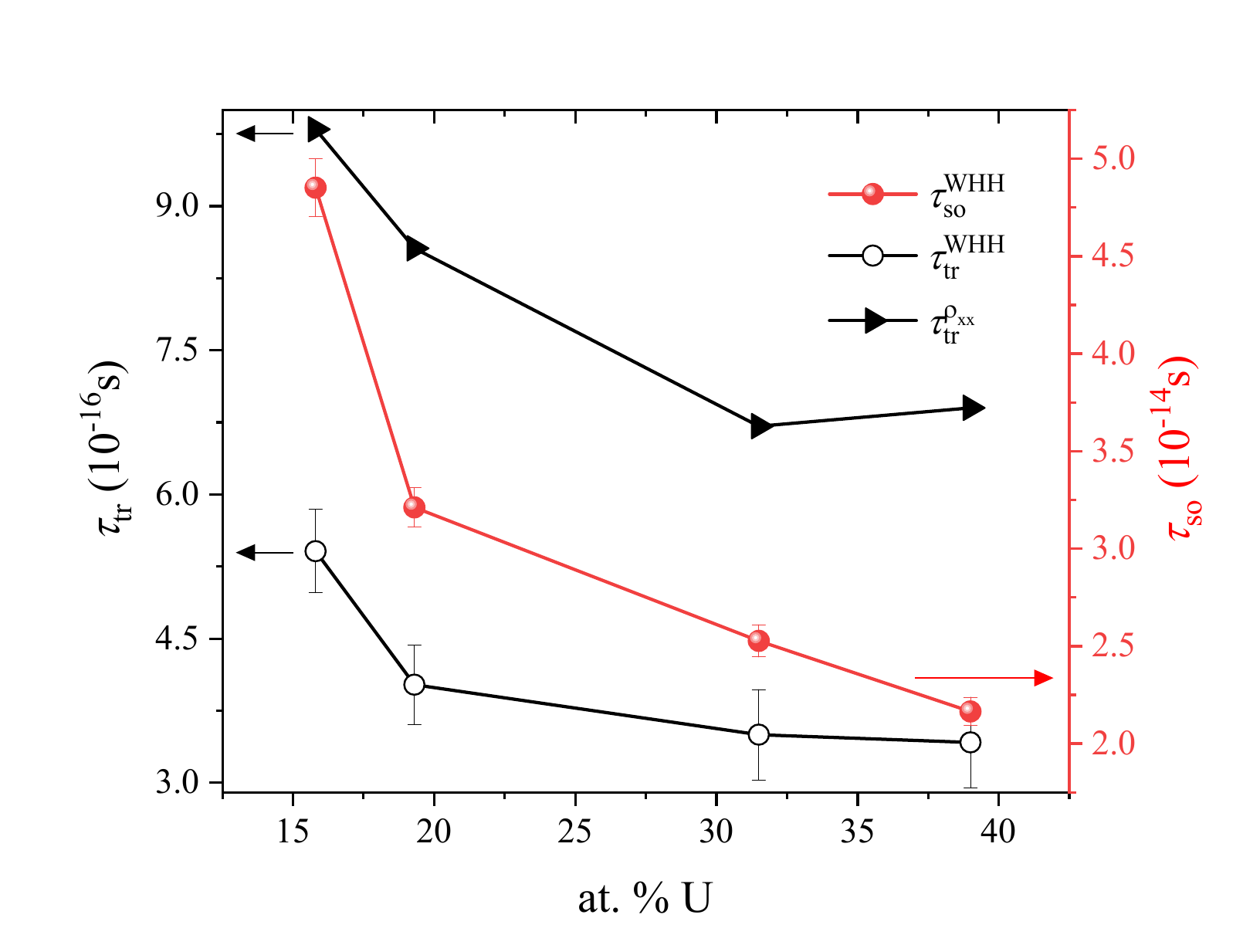}
    \caption{Various scattering times versus composition. Circles show the $ \tau_{\mathrm{tr}}^{\mathrm{WHH}}$ (open points) and $ \tau_{\mathrm{so}}^{\mathrm{WHH}}$ (closed points) times evaluated from the WHH fit to the $H_{\mathrm{c2}}(t)$ data, and  $\tau_{\mathrm{tr}}^{\mathrm{\rho_{\mathrm{xx}}}}$ values (triangles) were calculated from a single band Drude model. Lines are a guide to the eye.}
    \label{fig:ScatteringTimes}
\end{figure}

Additionally, the TRS time was also calculated using a simple single band Drude model based on the measured $\rho_{\mathrm{xx}}$, and assuming a band effective mass $m^{\ast}=m_{e}$ and the bulk literature Nb electron density, $n_{\mathrm{Nb}}=5.56\times10^{28}$ m$^{-3}$. The value of  $\tau_{\mathrm{tr}}^{\rho_{\mathrm{xx}}}$,  together with $ \tau_{\mathrm{tr}}^{\mathrm{WHH}}$ and $ \tau_{\mathrm{so}}^{\mathrm{WHH}}$ are shown in Figure \ref{fig:ScatteringTimes} as a function of at$.$ \% U, with the two values of $\tau_{\mathrm{tr}}$ agreeing within a factor of two, despite the simplified assumptions. A clear decrease in the values of the SOS time, and both measures of the TRS times with increasing at$.$ \% U is seen, suggesting an Elliott-Yafet-type mechanism \cite{kiss_elliott-yafet_2016, boross_unified_2013} where the spin-orbit scattering in the system is dominated by scattering due to the U atoms. This is further elaborated in the discussion section.

\section{Discussion}
\label{discussion}
Beginning with the x-ray characterisation data, the Nb (110) peak systematically shifted with increasing at$.$ \% U, consistent the Vegard's law expectations. No other measurable peaks were observed in the specular scan range, suggesting the U atoms entered the Nb {\it bcc} structure substitutionally. The U atoms therefore are located on sites with the same crystal symmetry as for the {\it bcc} $\gamma$-U system. The increase in electron density from both XRR and XRD observed provides additional evidence of the U atoms with more electrons entering the {\it bcc} structure of the host (Nb) cell. Previous studies have stabilised U in the {\it bcc} $\gamma$ structure in two main ways: splat cooling of higher temperature melts in alloys containing a range of different elements (e$.$g$.$ Mo, Nb, Pt, Zr, Ti, Ru), and epitaxial stabilisation combined with Mo substitution \cite{chaney_tuneable_2021}. Here, at lower U densities, we have achieved a similar structure for the U atoms, but in a more dilute limit, with a polycrystalline character.  

From the electronic transport and magnetotransport measurements, two main regions of temperature are of interest. The first region of interest is around the feature in resistivity that was observed for $4$ K $\leq {T} \leq 20 $ K and the second the superconducting region below $2$ K. We note more broadly that the absolute values of the resistivity and the form of the $\rho_{\mathrm{xx}}(T)$ data place these samples close to the Mooij limit for conductivity in highly disordered metals. We have confirmed that the Ge cap layer is highly resistive at these thicknesses and temperatures. Given the thickness of the Nb:U layers and the coherence length of the superconducting state we do not expect any influence of the cap material. Since the samples were deposited at room temperature we do not expect significant interdiffusion of Ge into the Nb:U layers here either.

\begin{figure}[h]
    \centering
    \includegraphics[width=1.0\linewidth]{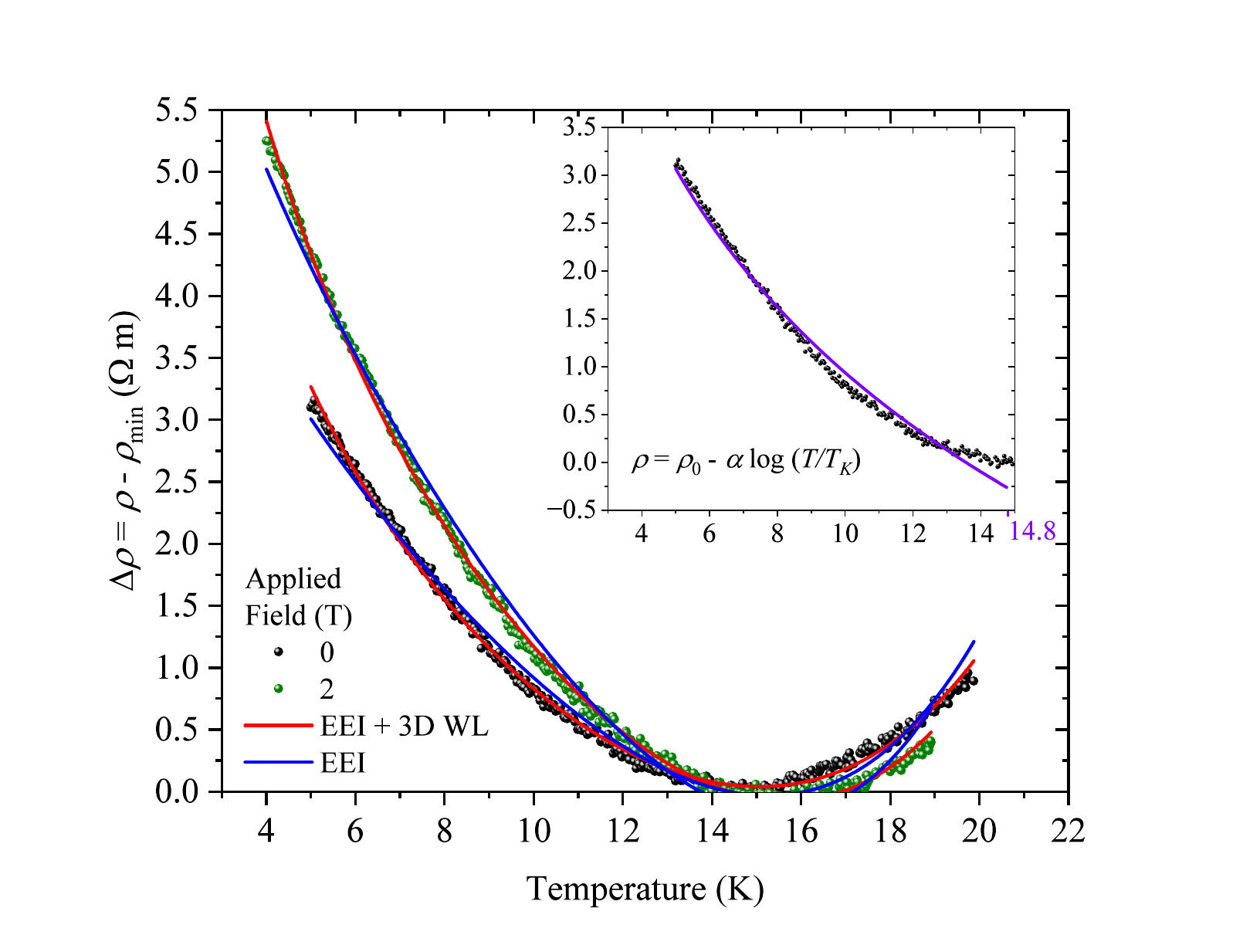}
    \caption{Change in resistivity with temperature measured from the resistivity at $T^{\star}$, $\rho_{\mathrm{min}}$, in two applied magnetic fields for the sample with 15.8 at$.$ \% U. Blue solid line is the best fit to the EEI only model. Red line is the best fit to a sum of the EEI and 3D WL forms. Inset: line is the best fit to the zero field resistivity data using the Kondo form $\rho = \rho_0 - \alpha\log(T/T_{\mathrm{K}})$.}
    \label{fig:LowTResistivityAppliedFields}
\end{figure}

Focussing first on the feature in the $\rho_{\mathrm{xx}}(T)$, which manifests as a local minimum for the two lowest resistivity samples. We refer to the temperature of the feature as $T^{\star}$. Such features in $\rho_{\mathrm{xx}}(T)$ data are typically attributed to three possible different effects: the Kondo effect \cite{kondo_resistance_1964}, electron-electron interactions (EEI) \cite{joseph_electronelectron_2020}, and localisation physics \cite{giordano_weak_2010}. In order to try to distinguish between these different possible origins, we have tracked magnetic field dependence of this effect. The magnetotransport measurements in this temperature range for sample with $15.8$ at$.$ \% U are shown in Figure \ref{fig:LowTResistivityAppliedFields}. We see that the resistivity at low temperatures shows a positive magnetoresistance, in contrast to the expected behaviour for the Kondo effect in the limit ${T}<{T}_{\mathrm{K}}$, where a resistivity decrease is expected with increasing applied fields due to the suppression of spin-flip scattering processes. However, it should be noted that the shift in $T^{\star}$ towards higher temperatures with applied fields in Figure \ref{fig:MinimaShift} might be explained in the context of competition between the Kondo energy scale $k_{\mathrm{B}}T_{\mathrm{K}}$ and the Zeeman energy, $\sim \mu_{\mathrm{B}}B$, which prevents the formation of a Kondo singlet via spin-flip scattering processes. The shift to higher temperatures reflects the higher energy needed to ``unlock" the impurity spins to participate in this process and compete against the Zeeman energy. Furthermore, in the inset of \ref{fig:LowTResistivityAppliedFields}, it can also be seen that the logarithmic fit for the Kondo effect, $\rho = \rho_0 - \alpha\log(T/T_{\mathrm{K}})$, where $T_{\mathrm{K}}$ is the Kondo temperature, does not fit the data well, especially at higher temperatures approaching the local minimum. 

\begin{figure}[h]
    \centering
    \includegraphics[width=1.0\linewidth]{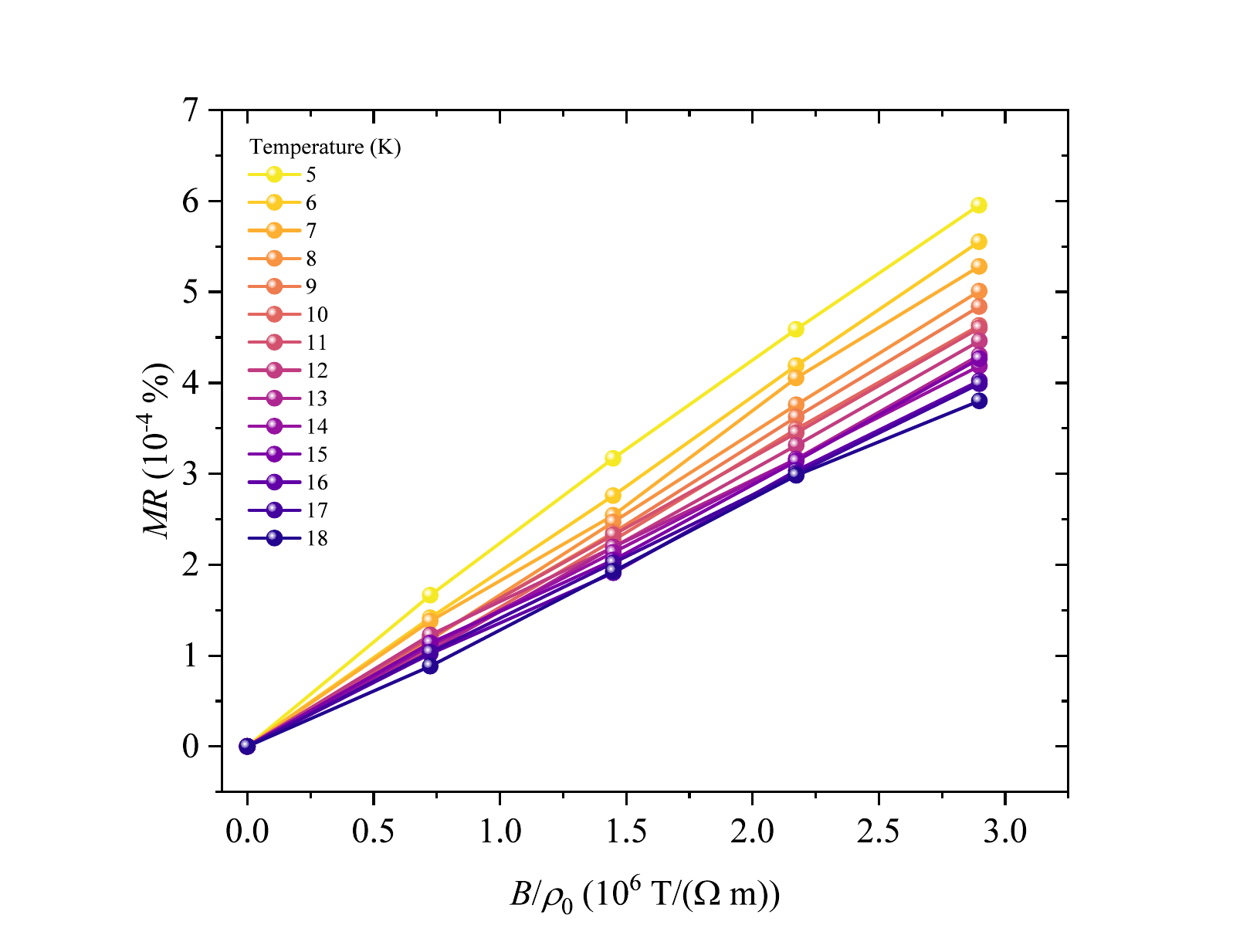}
    \caption{Spread of the magnetoresistance (MR) with applied fields normalised to the resistivity at zero field for temperatures in the range $5$ K $\leq T \leq 18$ K for the sample with 15.8 at$.$ \% U. }
    \label{fig:KohlerPlot}
\end{figure}

Attributing the observed dip in resistivity to the Kondo effect is unclear therefore, especially in a disordered system consisting of a heavy element with large SOC. We have therefore also considered two other possible sources of the resistivity feature: electron-electron interactions (EEI) and 3D weak localisation (WL). The EEI interaction was modelled using the functional form $ \rho = \rho_{0} + \alpha T^5 - \beta^{1/2} $ \cite{niu_evidence_2016}, and the WL effect modelled via $ \rho = \rho_{0} + \alpha T^5 - \beta^{1/2} + \lambda \ln(T)$, as shown by the lines in Fig$.$ \ref{fig:LowTResistivityAppliedFields}. Both fits were able to accurately model the resistivity data for $4$ K $\le T \le 20 $ K, therefore no firm conclusion about the origin of the feature at $T^{\star}$ can be drawn at this stage. We note that the Kohler's scaling for various temperatures in the same range - shown in Figure \ref{fig:KohlerPlot} - shows clear deviation from the expected scaling suggesting that an additional scattering mechanism may be activating at lower temperatures. How the feature in the $\rho(T)$ data - and the underlying physics causing it - may or may not be related to the suppression of $T_{\mathrm{c}}$ and $H_{\mathrm{c2}}(0)$ with increasing at$.$ \% U is unclear at this time. Potentially both magnetic pair-breaking and/or disorder-induced scattering may be contributing factors. 

From the WHH fits to the superconducting data, it can be seen in Table \ref{tab:Hc-table-summary} that $\mu_{0}H_{\mathrm{c2}}^{\mathrm{WHH}}(\mathrm{T=0})$ are all below the BCS Pauli limit, $\mu_{0}H_{\mathrm{P}}^{\mathrm{BCS}}$ calculated assuming a pure singlet state for the Cooper pairs. Notably however the ratio $H_{\mathrm{P}}^{\mathrm{BCS}}/H_{c2}^{\mathrm{WHH}}$ is much closer to unity at larger at$.$ \% of U, suggesting that the SOC strength increases as more U is added to the system. The variation of the Maki parameter with at$.$ \% U also indicates a shift from mainly orbital depairing dominant to more Pauli-limited samples. 

The SOS parameter $\lambda_{\mathrm{so}}$ was also extracted from the WHH fits, which allowed the scaling  of $\tau_{\mathrm{tr}}^{\mathrm{WHH}}$ and $\tau_{\mathrm{so}}^{\mathrm{WHH}}$ with at$.$ \% U to be determined. Both $\tau_{\mathrm{tr}}^{\mathrm{WHH}}$ and  $\tau_{\mathrm{so}}^{\mathrm{WHH}}$ and decreased with increasing at$.$ \% U, as shown in Figure \ref{fig:ScatteringTimes}, consistent with the Elliott-Yafet (EY) \cite{elliott_theory_1954, yafet_g_1963} mechanism of spin relaxation. In this EY picture, spin relaxation is linked with momentum scattering at the U impurity atoms with large SOC. We note that in principle the changes in crystallite size discussed above can also impact the momentum scattering rate, and therefore the scaling observed. However the XRD data suggested in relatively small {\it increase} in crystallite size with increasing U content, which would be expected to alter the mean-free path in the opposite way to the expected EY scaling. We conclude therefore that the crystallite size change is a minor effect here.   

Although WHH theory could be well fitted to the data, it is important to note that this WHH model assumes a single band, while the superconductivity in Nb is inherently multi-band \cite{saunderson_gap_2020}. With the additional complexity of the $5f$ electrons from the U atoms into this system, a full calculation of the change of the chemical potential, Fermi surface size and shape (and therefore average Fermi velocity) with doping is required to fully quantitatively understand the physics. However we emphasise that the high U concentrations used here give rise to significant scattering and short electronic mean-free paths, meaning that there is relatively large inter- and intra-band blurring. This will tend to wash out multiband features in the $H_{c2}(T)$, for example. Indeed we have seen no features suggesting multiband physics in the superconducting data.   

\section{Summary and conclusions}
\label{conclusions}
Thin films of $\mathrm{Nb}_{1-x}\mathrm{U}_{x}$ solid solutions with at$.$ \% U concentration between 15 - 40\% have been prepared via d$.$c$.$ magnetron sputtering at room temperature. X-ray characterisation of the samples revealed a systematic shift of the (110) Nb peak with U concentration causing an increase in the {\it bcc} lattice parameter and the electron density observed within the system. 

Resistivity and magnetoresistance measurements revealed a possible complex interplay of electron-electron interaction and localisation physics between $4$ K $\leq T \leq 30$ K leading to an observed feature in the resistivity sensitive to both U concentration and applied field. Kohler's scaling was not followed in the range $5$ K $ \le T \le 18$ K \cite{scaling_comment}. 
Superconductivity was observed in all samples below $2$K. The analysis of the superconducting region revealed a proportional relationship between spin-orbit scattering and transport scattering times, $\tau_{\mathrm{so}}\propto\tau_{\mathrm{tr}}$ confirming an Elliott-Yafet-type spin relaxation in the system. Given the the coexistence of superconductivity and strong SOC in these samples, we emphasise that these types of samples - incorporating heavy actinide elements with more conventional superconducting materials - open up a significant range of possible future directions. These include in the fields of superconducting spintronics - for example the generation of odd frequency triplet states - and topological superconductivity.

\section*{Acknowledgements}
We acknowledge the funding and support from the UKRI Engineering and Physical Sciences Research Council, in particular this work utilised the National Nuclear User Facility FaRMS, grant: EP/V035495/1. KB and LO received support from the School of Physics, University of Bristol, as part of their undergraduate studies.

\newpage
\section*{References}

\bibliographystyle{unsrt} 
\bibliography{finalbib}

\end{document}